\documentstyle[12pt,epsf]{article}

\newcommand{\beq}{\begin{equation}}
\newcommand{\eeq}{\end{equation}}
\newcommand{\bea}{\begin{eqnarray}}
\newcommand{\eea}{\end{eqnarray}}
\newcommand{\bc}{\begin{center}}
\newcommand{\ec}{\end{center}}

\begin{document}
\title{Tunneling Ionization Rates from Arbitrary Potential Wells}
\author{{\normalsize G. N. Gibson$^1$, G. Dunne$^1$, and K. J. Bergquist$^2$}\\
{\normalsize $^1$Department of Physics, University of Connecticut, Storrs,
CT 06269}\\
{\normalsize $^2$Department of Physics, Wake Forest University, Winston-Salem, NC 27109}}
\date{}
\maketitle

\begin{abstract}
We present a practical numerical technique for calculating tunneling
ionization rates from arbitrary 1-D potential wells in the presence of a
linear external potential by determining the widths of the resonances in the
spectral density, $\rho(E)$, adiabatically connected to the field-free bound
states. While this technique applies to more general external potentials, we
focus on the ionization of electrons from atoms and molecules by DC electric
fields, as this has an important and immediate impact on the understanding
of the multiphoton ionization of molecules in strong laser fields.
\end{abstract}


While the behavior of atoms and molecules in strong laser fields has been
studied for almost two decades, new phenomena continue to arise and a
unified picture for either does not yet exist \cite{WAL94,HAN97,TAL97}.
However, simple models for tunneling ionization \cite{AMM86,COR89} have
provided a crucial baseline for understanding atoms in strong fields, as
deviations from this basic process in ion yields and electron energy spectra
have pointed to new physics \cite{WAL94,HAN97,COR93}. Conversely, the lack
of a general tunneling model for molecules has seriously impeded the
interpretation of molecular data which clearly hints at even richer
phenomena \cite{TAL97,SEI95} than for atoms due to the extra degrees of
freedom present in a molecule.

Ionization from even diatomic potential wells are more difficult to treat
than atoms because the barrier to ionization can shift between the barrier
separating the two wells and the outer barrier as a function of field
strength and internuclear separation \cite{COD89} (see Fig.~1). These
situations are quite distinct and, indeed, this competition between barriers
leads to interesting and complex behavior. Thus, it is critical for any
tunneling ionization model to work for arbitrary potential wells so as to
treat these different situations uniformly. Furthermore, a useful model for
molecules in strong laser fields must be non-perturbative in both the rates
and the Stark shifts. In this Letter we introduce a practical
non-perturbative technique for computing tunneling ionization rates from an
arbitrary 1-D potential well. After outlining the general method, we apply
it to a single-well (`atomic') potential, yielding excellent agreement with
existing theoretical analyses. Then we apply our method to a double-well
(`molecular') potential, like that shown in Fig. 1, to illustrate the
generality of the method and to show new features which arise due to the
double-well structure. However, we emphasize that the method is exactly the
same for any arbitrary shaped 1-D potential well.

The problem of determining ionization rates and Stark shifts
non-perturbatively for an arbitrary potential in 1-D can be reduced to
calculating the spectral density, $\rho(E)$. For potentials with strictly
bound states $\rho(E)$ becomes a series of delta functions at the energy
eigenvalues. Numerically integrating the Schr\"odinger equation (SE)
determines these energies by finding the energy at which the wavefunction
satisfies the given boundary conditions. This simple approach cannot be
extended to the question of tunneling ionization because, by definition, the
electron can escape to infinity. Given this fact, the electron's state lies
in the continuum; hence, all energies are allowed and simply integrating the
SE at a particular energy provides little information, on its own. Rather,
one must return to calculating $\rho(E)$ in general. In the presence of a
linear potential, the delta functions in $\rho(E)$ for the bound states will
evolve into Lorentzian-like resonances whose center positions reflect the
Stark shift of the states and widths give their decay rates. As decay of the
levels can only occur through ionization the width then directly gives the
tunneling rate.

The limitation to 1-D potential wells may seem overly restrictive and
requires some discussion. It is well known that ionization rates of atoms
and molecules depend strongly on field strength \cite{AMM86}. For this
reason ionization occurs overwhelmingly in the direction of the electric
field, essentially reducing the problem to 1-D. For diatomic molecules, the
ionization rate is much greater when the field is along the internuclear
axis \cite{DIE93} and, again, reduces the problem to 1-D.

While a full discussion of $\rho(E)$ is rarely presented in standard books
on quantum mechanics, a general approach exists for calculating $\rho(E)$,
based on the Weyl-Titchmarsh-Kodaira (WTK) spectral theorem \cite{TIT46}.
Remarkably, this approach is extremely well suited to numerical
implementation. Essentially, the spectral theorem states that the
``resolution of the identity'' can be expressed as 
\begin{equation}
\delta\left(x-y\right)=\sum_{i,j=1,2}\int^\infty_{-\infty} dE\,
\phi_i\left(x,E\right)\phi_j\left(y,E\right)\rho_{ij}\left(E\right)
\label{SPEC-THRM}
\end{equation}
where $\phi_{1,2}(x,E)$ are two linearly independent eigenfunctions of the
SE at energy $E$ and $\rho_{ij}(E)$ is the spectral density matrix. Before
discussing how to calculate and interpret $\rho_{ij}(E)$, we must consider
the question of normalization, as, in general, some of the states may lie in
the continuum.

The usual concept of a finite square integral clearly cannot be used for the
normalization of continuum wavefunctions. In the WTK approach, normalization
is imposed locally by setting the Wronskian of the eigenfunctions equal to
one which also ensures that $\phi_1$ and $\phi_2$ are linearly independent.
Thus, we can choose the following initial conditions at any point $x=c\,$: 
\begin{equation}
\begin{array}{cccc}
\phi_1\left(c\right)=1, & \phi^{\prime}_1\left(c\right)=0, & 
\phi_2\left(c\right)=0, & \phi^{\prime}_2\left(c\right)=1.
\end{array}
\label{BOUND-COND}
\end{equation}
For convenience, we generally choose $x=0$. The complete functions $%
\phi_i(x) $ can now be found by numerically integrating the SE away from the
origin using these initial conditions.

To construct the spectral density matrix $\rho_{ij}(E)$, we first find the
linear combinations of $\phi_1$ and $\phi_2$, 
\begin{equation}
\psi_\pm\left(x,\lambda\right)=\phi_1\left(x,\lambda\right)
+m_\pm\left(\lambda\right)\phi_2,
\end{equation}
such that $\int_{-\infty}^0 dx\,\left|\psi_-\right|^2$ and $\int_0^\infty
dx\,\left|\psi_+\right|^2$ are both finite, where $\lambda=E+i\varepsilon$
and $\varepsilon\geq 0$. In other words, when the energy has a positive
imaginary part, $\psi_\pm\left(x,\lambda\right)$ must decay sufficiently
rapidly as $x\rightarrow\pm\infty$ to have a finite square integral on their
respective half-line. This defines the functions $m_\pm\left(\lambda\right)$%
. The spectral density matrix is then \cite{TIT46}: 
\begin{equation}
\rho_{ij}(E)={\frac{1 }{\pi}}\lim_{\varepsilon\rightarrow 0}{\rm Im} {\frac{%
1 }{m_--m_+}} \left[ 
\begin{array}{cc}
1 & {\frac{m_-+m_+ }{2}} \\ 
{\frac{m_-+m_+ }{2}} & m_-m_+
\end{array}
\right].  \label{SPEC-DEN}
\end{equation}
If the asymptotic wavefunctions of the external potential are known
analytically, as is often the case, all that is needed to evaluate $m_\pm$
is the logarithmic derivative of the correctly decaying asymptotic
wavefunctions of just the external potential, $\theta_\pm\left(x,\lambda%
\right)$: $f_\pm\left(x,\lambda\right)
=\theta^{\prime}_\pm\left(x,\lambda\right)/\theta_\pm\left(x,\lambda\right)$ 
\cite{AIRY-NOTE}. We can then solve for $m_\pm\left(\lambda\right)$: 
\begin{equation}
m_\pm\left(\lambda\right)=-\left({\frac{ \phi^{\prime}_1-f_\pm\phi_1 }{%
\phi^{\prime}_2-f_\pm\phi_2}}\right).  \label{M-PM}
\end{equation}
Note that the overall normalization of the asymptotic wavefunctions is not
important, as the logarithmic derivative is sufficient to determine $%
m_\pm\left(\lambda\right)$.

This entire procedure is very straightforward to implement numerically:
starting with the boundary conditions in Eq.~(\ref{BOUND-COND}), $\phi_1$
and $\phi_2$ are found by integrating the SE away from the origin with a
standard Runge-Kutta technique at a particular complex energy, $\lambda$. At
the same time, $m_\pm$ are evaluated from Eq.~(\ref{M-PM}). Ideally, one
would integrate out to $\pm\infty$ to evaluate $m_\pm\left(\lambda\right)$
exactly. Since this is not possible numerically, we simply evaluate $%
m_\pm\left(\lambda\right)$ as a function of $x$ far enough from the local
potential to where it converges to a constant value. At this point, $%
m_\pm\left(\lambda\right)$ have been determined as well as $\rho_{ij}(E)$
from Eq.~(\ref{SPEC-DEN}).

To finish the problem, we must interpret the matrix in Eq.~(\ref{SPEC-DEN}).
The spectral density matrix, $\rho_{ij}(E)$, is positive definite \cite
{TIT46} and, thus, its eigenvalues $\Lambda_1$ and $\Lambda_2$ are positive
and represent scalar spectral density functions. In general, there are two
functions because of the possibility of a two-fold degeneracy
(corresponding, for example, to left and right moving solutions) on the
whole line. However, in the case of a linear external potential (Fig.~1),
one direction is inaccessible to a free particle and no degeneracy exists.
In this case, either $\Lambda_1$ or $\Lambda_2$ will be on the order of $%
\varepsilon$ and is unimportant, leaving just one eigenvalue, resulting in a
simple scalar spectral density. For problems on the half-line, $x\geq 0$,
with a physical boundary condition at $x=0$, the spectral density matrix
also reduces to a single spectral density function \cite{TIT46,DEA82}.
Finally, we must consider taking the limit $\varepsilon\rightarrow 0$.
Again, numerically, this is simple to specify: if the state can tunnel, it
will have a ``natural'' width $\Gamma$ giving the tunneling rate and an
``artificial'' width determined by $\varepsilon$. One then decreases $%
\varepsilon$ until $\Gamma\gg\varepsilon$. We have noticed that the
numerical results are surprisingly robust to changes in the various
numerical parameters. We attribute this to the local specification of the
normalization and initial conditions for the numerical integration and the
ease in determining the $m_\pm$ values.

In order to test this method we considered tunneling ionization of a
hydrogen atom by a DC electric field: 
\begin{equation}
V(x)={\frac{1}{\sqrt{x^{2}+a^{2}}}}-Fx,
\end{equation}
where $F$ is the field strength and $a=\sqrt{2}$ for a hydrogen atom. This
1-D form of the Coulomb potential \cite{JAV88} has proven to be an excellent
model for strong field calculations because it preserves the long range $1/r$
dependence and has an infinite Rydberg-like series of bound states while
removing the catastrophic singularity that occurs for 1-D. Furthermore, it
has been used recently in direct calculations of the time-dependent
Schrodinger equation for molecules \cite{SEI95}. Fig.~2 shows the spectral
function at a moderate field strength. The first two low lying states in the
Rydberg series can be seen, while the higher states have broadened and
merged together. We stress that the blurring of the higher states is a real
physical effect, due to the lowering of the potential barrier by the
external field, and is not a limitation of the method. In a weaker field
more of the excited states would be seen as well-defined peaks. Fig.~3 shows
a typical unbound wavefunction at the peak of a resonance. At this point,
most of the wavefunction is ''inside'' the potential well, although it is
still a continuum state. Fig.~4 shows the DC Stark shift as well as the
first term in the second order perturbation theory expansion for the 1-D
potential. Finally, Fig.~5 shows the tunneling ionization rate of the ground
state as a function of field strength. In order to compare to a 3-D
potential, we performed an angular average of the 1-D result: 
\begin{equation}
\Gamma (F)={\frac{1}{2\pi }}\int d\Omega \,\Gamma \left( F\cos \theta
\right) .
\end{equation}

As seen in Fig. 5,
this averaged result agrees quite well with the standard 3-D static atomic
tunneling model of ADK \cite{AMM86,COR89}. However, the slight discrepancy
at the higher intensity is
intriguing. The ADK model depends on the ionization potential and the
field strength, but does not take into account the DC Stark shift.
While the shifts are small, ionization rates depend exponentially on
ionization potential and the effect of the Stark shift on the
ionization rate will increase with
intensity. We recalculated the ADK rate using our calculation of
the DC Stark shift, and, as seen in Fig. 5 the agreement
between our angularly averaged model and the ADK result including the
DC Stark shift is excellent.

The 'molecular' or double-well potential has a much richer spectrum in an
external field than an atom, and we cannot give a comprehensive treatment,
here. However, as an example, we show, in Fig. 6, the effect of the
separation of the double-well potential on ionization rates for both the
ground and first excited states. The potential consists of two 1-D Coulomb
potentials (Eq. 6) separated by a distance R, and the external field, as in
Fig. 1. Using exactly the same method of calculating the spectral function
and finding the widths of the peaks associated with the ground and first
excited states, we find that the ionization rate is strongly dependent on
the well separation, becoming greater at larger separations. Moreover, the
qualitative shape of the ionization rate vs. intensity curve depends on the
separation with apparently no saturation setting in for the excited state at
the larger separation. In fact, these are just the conditions for electron
localization and enhanced ionization predicted in Ref. \cite{SEI95}. These
ionization rate curves can provide a framework to analyze experimental ion
yield data from molecules.

In summary, we have described a practical numerical technique for
calculating spectral density functions for arbitrary 1-D potentials on the
whole line from which we can extract level shifts and tunneling rates. This
will be of great value for interpreting data from atoms and molecules in
strong laser fields.

\vskip 1cm

We thank S.~A.~Fulling and W.~P.~Reinhardt for helpful discussions. We would
like to acknowledge support from the NSF under grant PHY-9502935. G.~N.~G.
was also supported through funding as a Cottrell Scholar of Research
Corporation. K.~J.~B. was supported by the Research Experience for
Undergraduates (REU) program of the NSF. G.~D. acknowledges support from the
DOE under grant DE-FG02-92ER40716.00.


\newpage

\begin{center}
Figure Captions
\end{center}

\begin{enumerate}
\item  Molecular potential wells showing the importance of (a) the outer
barrier and (b) the inner barrier.

\item  Spectral density function for a model atom in an external field.

\item  Potential in an external field with wavefunction on resonance.

\item  Comparison of the 2nd order Stark shift calculation with perturbation
theory, in atomic units.

\item  Comparison of the angularly-averaged WTK tunneling rate model with a
3-D calculation (ADK) from Ref.~\cite{AMM86}.

\item  Ionization rate of the ground and first excited state of a
double-well potential for two different separations.
\end{enumerate}

\begin{figure}[h]
\epsffile{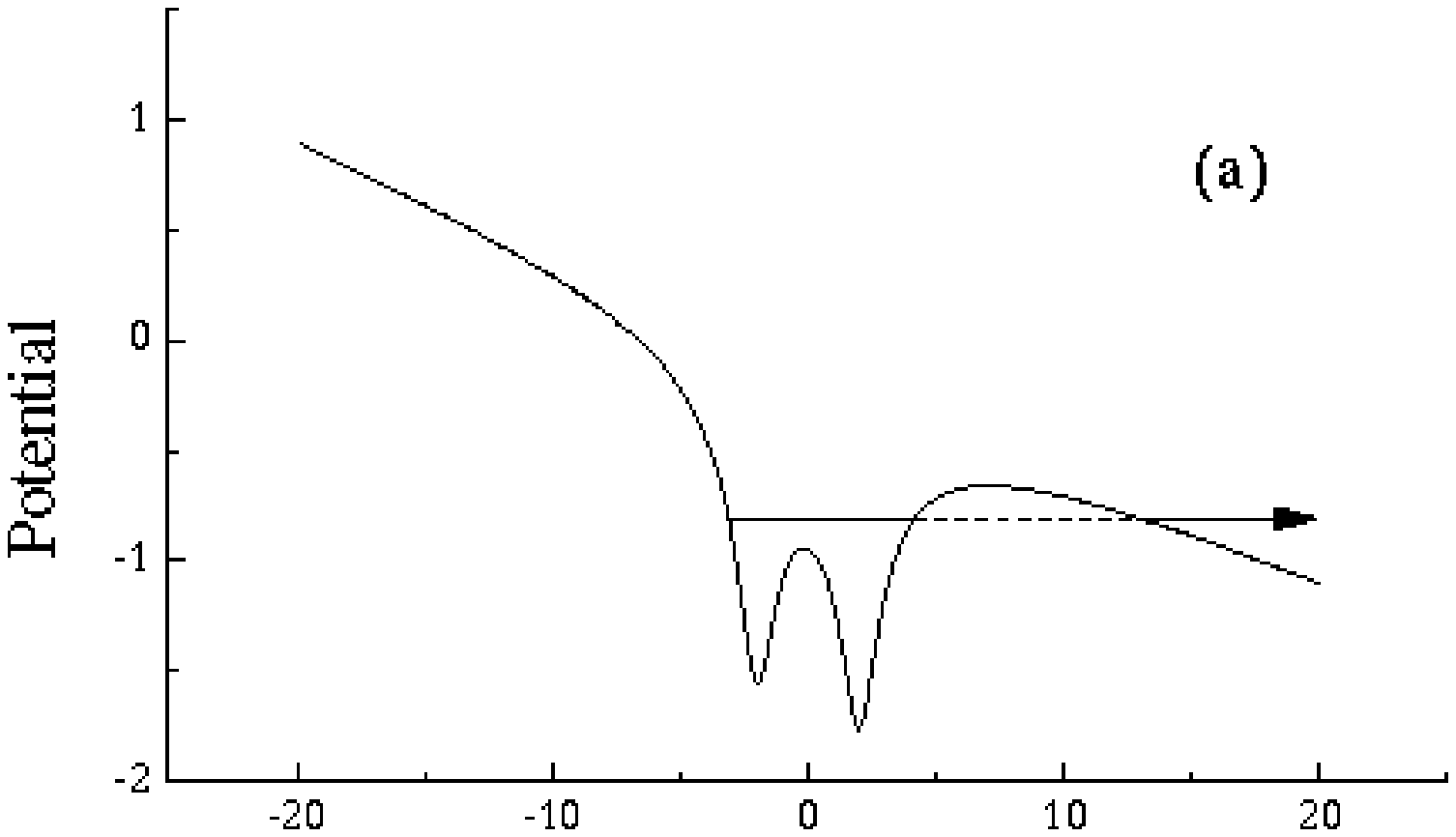}
\epsffile{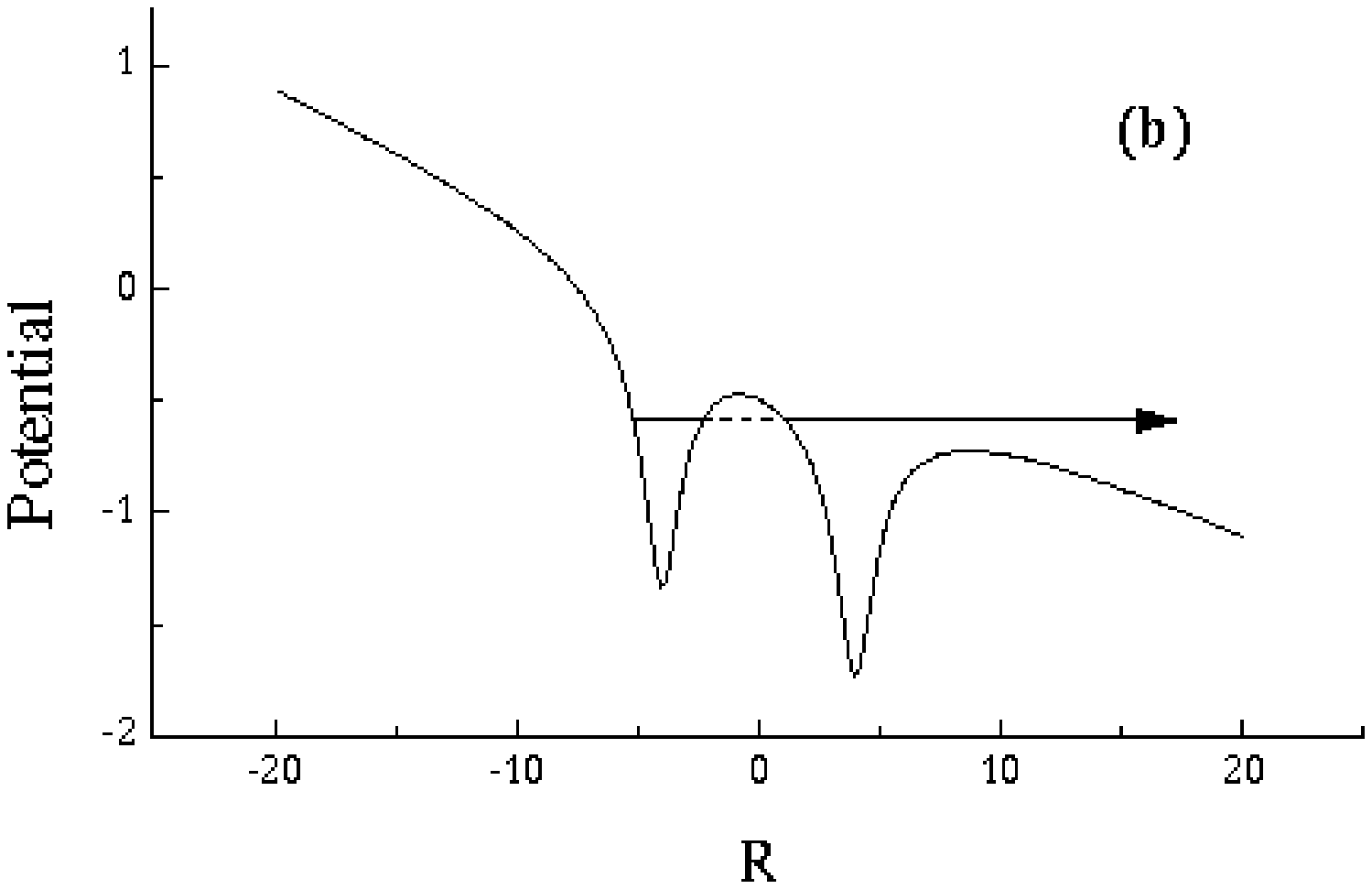}
\caption{Molecular potential wells showing the importance of (a) the outer
barrier and (b) the inner barrier.}
\label{f1}
\end{figure}

\begin{figure}[h]
\epsffile{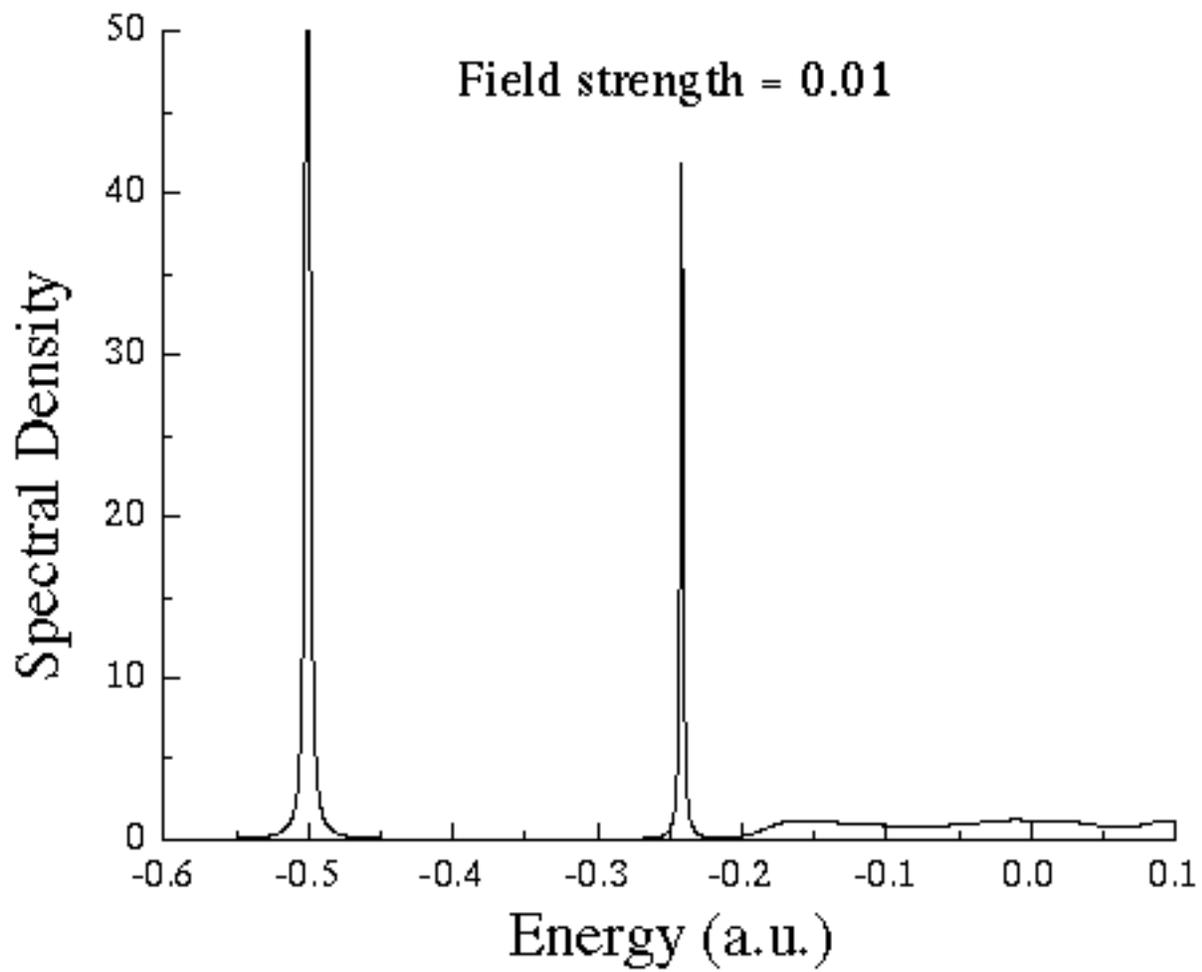}
\caption{Spectral density function for a model atom in an external field.}
\label{f2}
\end{figure}

\begin{figure}[h]
\epsffile{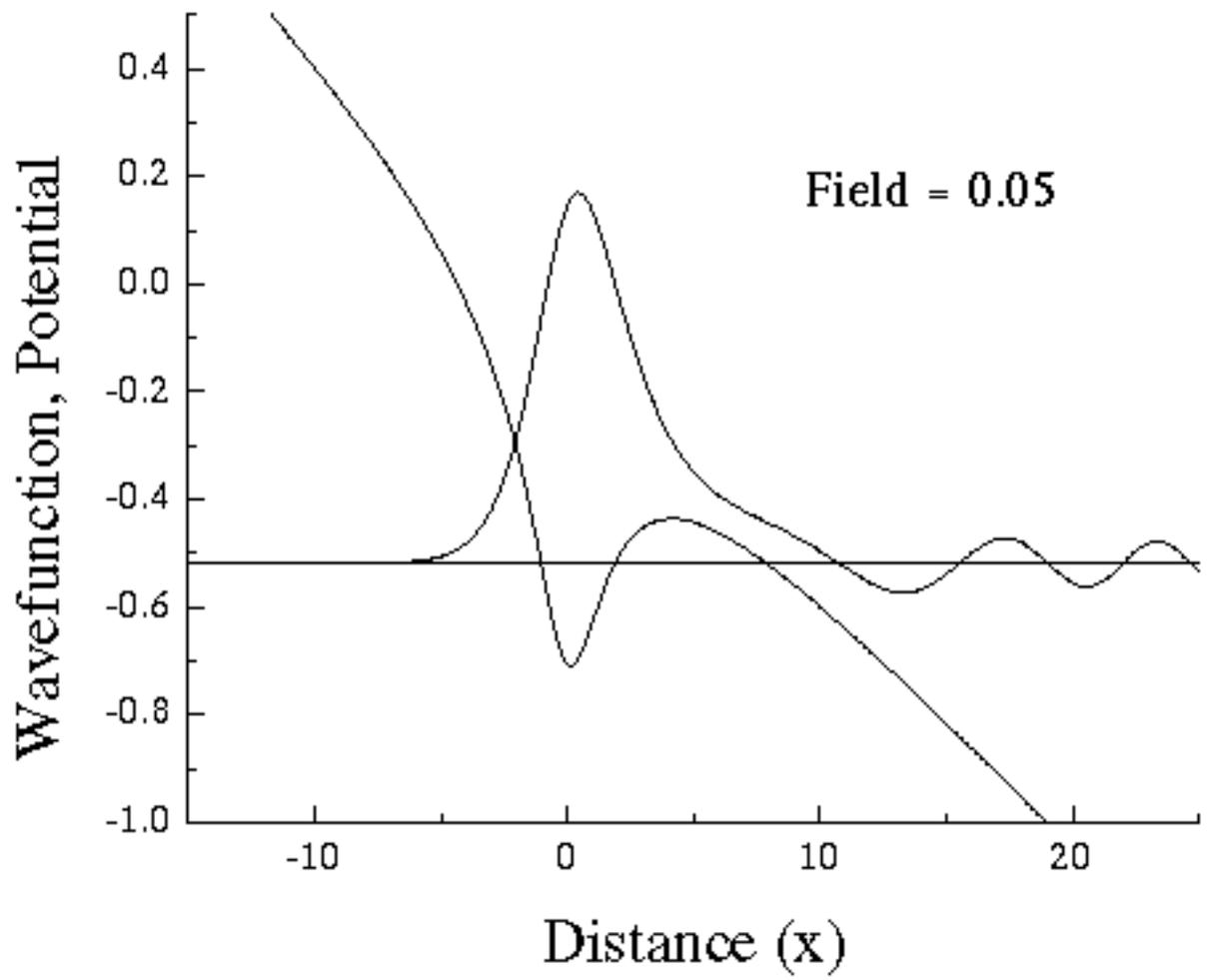}
\caption{Potential in an external field with wavefunction on resonance.}
\label{f3}
\end{figure}

\begin{figure}[h]
\epsffile{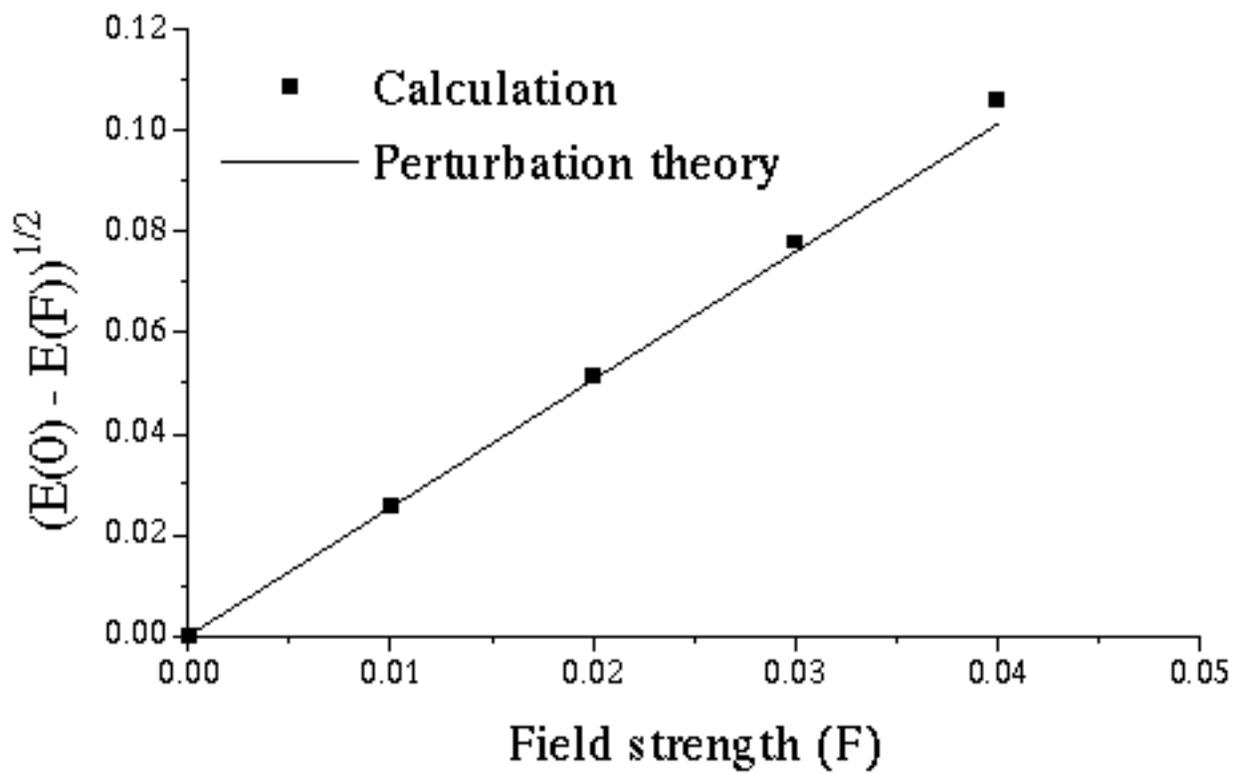}
\caption{Comparison of the 2nd order Stark shift calculation with perturbation
theory, in atomic units.}
\label{f4}
\end{figure}

\begin{figure}[h]
\epsffile{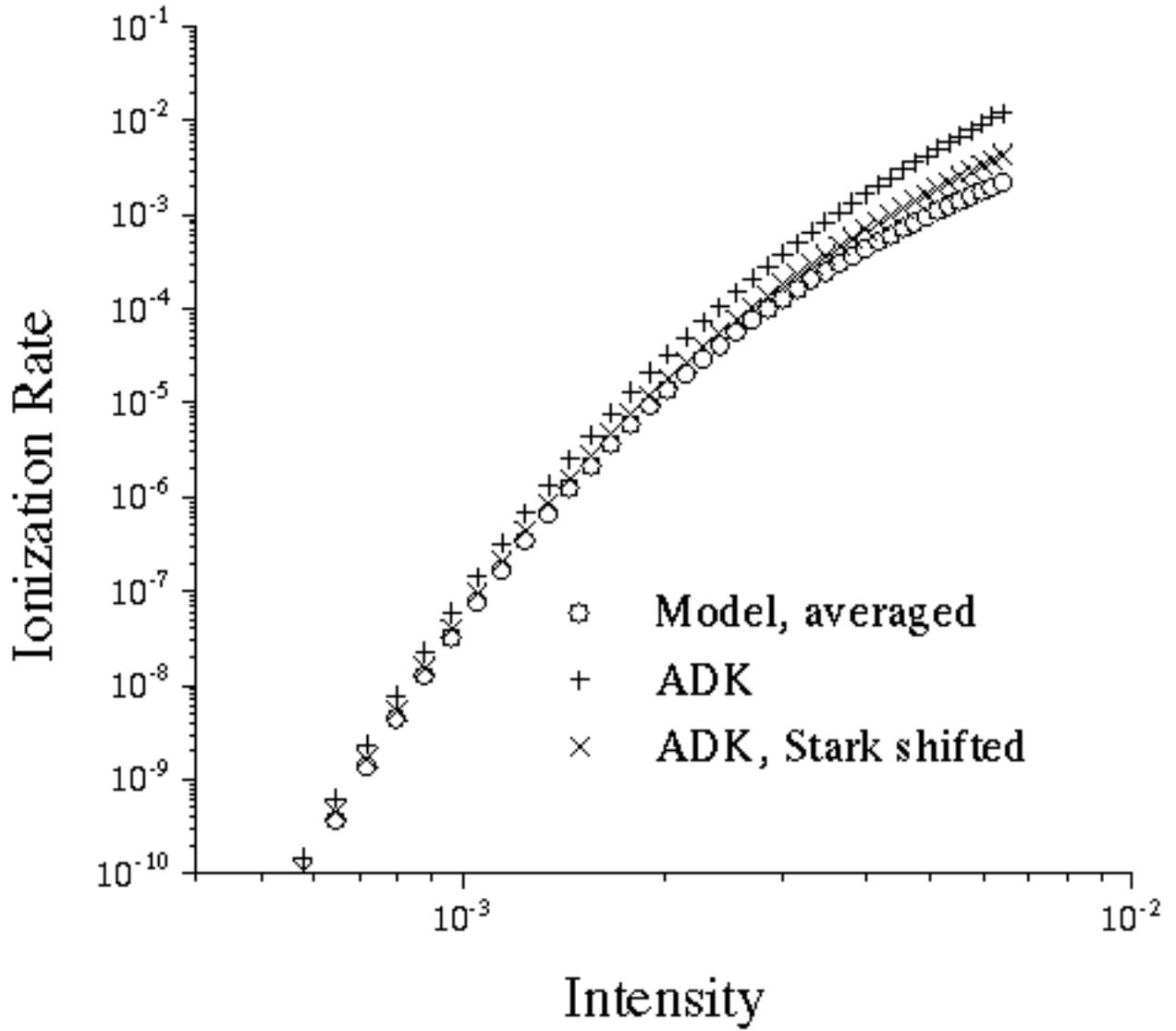}
\caption{Comparison of the angularly-averaged WTK tunneling rate model with a
3-D calculation (ADK) from Ref.~\protect{\cite{AMM86}}.}
\label{f5}
\end{figure}

\begin{figure}[h]
\epsffile{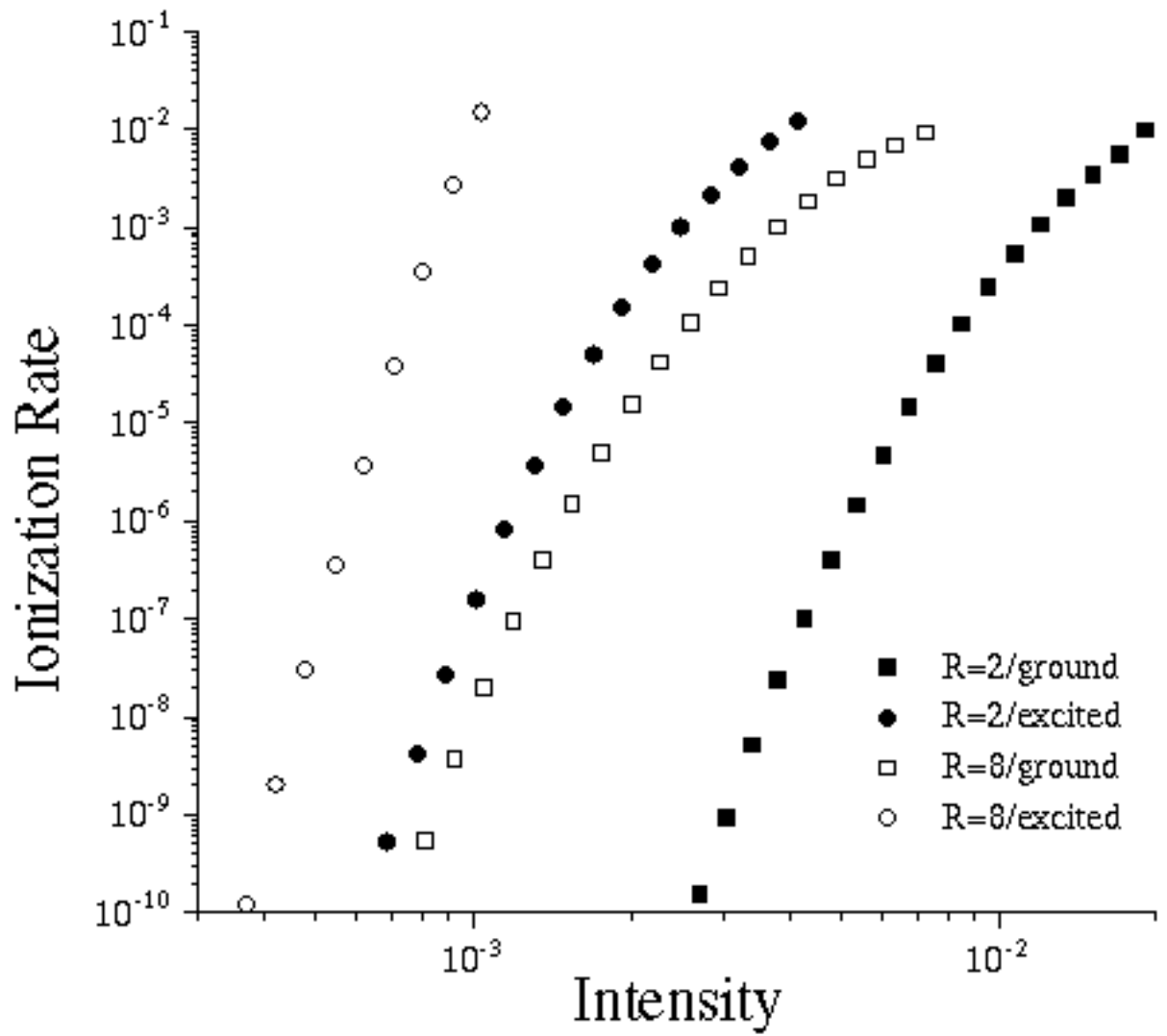}
\caption{Ionization rate of the ground and first excited state of a
double-well potential for two different separations.}
\label{f6}
\end{figure}


\begin{thebibliography}{99}
\bibitem{AIRY-NOTE}  For a linear potential $V(x)=-Fx$, the asymptotic
solutions are Airy functions, so 
\begin{eqnarray}
f_{1}(y) &=&-{\frac{\varepsilon }{2\sqrt{y}}}-{\frac{1}{4\left( y+\tilde{E}%
\right) }}+\left( \sqrt{y}+{\frac{\varepsilon }{2\sqrt{y}}}\right)  \nonumber
\\
f_{2}(y) &=&-{\frac{1}{4\left( y+\tilde{E}\right) }}+\sqrt{-y-\tilde{E}}, 
\nonumber
\end{eqnarray}
where $y=\left( 2F\right) ^{1/3}x$ and $\tilde{E}=\left( 2/F^{\,2}\right)
^{1/3}E$.
\end{thebibliography}
\end{document}